\documentclass[aps,pra,twocolumn,groupedaddress,showpacs]{revtex4}

\usepackage{graphicx}
\usepackage{amsmath}
\usepackage{amssymb}
\usepackage{bm}

\begin{document}
\title{Observation of deviations from ideal gas thermodynamics in a trapped Bose-Einstein condensed gas}
\author{F.\ Gerbier}
\altaffiliation[Present adress:]{Institut f{\"ur} Physik, Staudinger Weg 7, D-55128 Mainz; e-mail: gerbier@uni-mainz.de}
\author{J.\ H.\ Thywissen}
\altaffiliation{Department of Physics, University of Toronto,
Toronto, ON, M5S 1A7, Canada.}
\author{S.\ Richard}
\author{M.\ Hugbart}
\author{P.\ Bouyer}
\author{A. Aspect}
\affiliation{Groupe d'Optique
Atomique, Laboratoire
Charles Fabry de l'Institut d'Optique\footnote{UMR 8501 du CNRS},
91403 Orsay Cedex, France}
\date{\today}
\begin{abstract}
We have investigated experimentally the finite-temperature
properties of a Bose-Einstein condensed cloud of $^{87}$Rb atoms
in a harmonic trap. Focusing primarily on condensed fraction and
expansion energy, we measure unambiguous deviations from ideal-gas
thermodynamics, and obtain good agreement with a Hartree-Fock
description of the mixed cloud. Our results offer for the first
time clear evidence of the mutual interaction between the
condensed and thermal components. To probe the low-temperature
region unaccessible to the usual time-of-flight technique, we use
coherent Bragg scattering as a filtering technique for the
condensate.  This allows us to separate spatially the condensed
and normal components in time of flight, and to measure reliably
temperatures as low as $0.2~T_{\rm c}^0$ and thermal fractions as
low as 10\%.Finally, we observe evidence for the
limitations of the usual image analysis procedure, pointing out to
the need for a more elaborate model of the expansion of the mixed
cloud.
\end{abstract}
 \maketitle
%
%
\section{Introduction}

Trapped, dilute atomic gases offer a new opportunity to study the
interplay between quantum-statistical phenomena and interactions
in Bose systems \cite{nature_review}. A third ingredient, the
external trapping potential, also plays a key role in
understanding the properties of these gases \cite{dalfovo1999a}.
At finite temperatures, it leads to spatially distinct condensed
and thermal phases, a new behavior when compared to bulk quantum
fluids, in which both components overlap everywhere. This spatial
separation allows in particular a clear identification of the
condensed fraction through absorption imaging, in stark contrast
with superfluid $^4$He, where condensed fraction measurements are
only indirect \cite{sokol1993a}.

Much work has been devoted to the properties of condensed gases at
very low temperatures (much smaller than the critical temperature
$T_{\rm c}$), where the non-condensed fraction is negligible.
Then, in the so-called Thomas-Fermi (TF) regime
\cite{dalfovo1999a}, the static and dynamic behavior of the
condensate is essentially determined by the interplay between the
trapping potential and the atomic interactions. At higher
temperatures, less than but comparable to $T_{\rm c}$, a
significant thermal component is also present, typically much more
dilute than the condensate. In this situation, the kinetic energy
per thermal atom is larger than the mean-field energy, and
deviations from ideal gas behavior are small
\cite{ensher1996a,mewes1996a,marago2001a,han1998a,anderson1999a,schreck2001a}.
Although several key thermodynamical properties, such as condensed
fraction and average energy, are readily measurable
experimentally, a detailed comparison of experiments with
finite-temperature theories of the interacting cloud is to our
knowledge still lacking.

The goal of this paper is to contribute to fill this gap, by
investigating experimentally the thermodynamics of a trapped
$^{87}$Rb Bose gas below $T_{\rm c}$, where both a condensed and a
thermal component are present. We focus on two quantities, the
condensed fraction and the expansion (kinetic plus mean-field)
energy in the radial direction. We find that the condensed
fraction $N_0/N$ is significantly reduced with respect to the
ideal-gas law, $N_0/N=1-(T/T_{\rm c}^0)^3$, where $T_{\rm c}^0$ is
the ideal gas condensation temperature in the thermodynamic limit,
and that the expansion energy is increased, even for the thermal
atoms.

Our measurements thus clearly exclude ideal-gas behavior, and to
assess the importance of interactions, we compare them with two
mean-field theories of the interacting cloud that assume a
condensate in the TF regime. The simplest one is the so-called
``semi-ideal'' model \cite{minguzzi1997a,naraschewski1998a}, which
considers the thermal cloud as a quantum-saturated gas evolving in
the combined trapping plus condensate-mean-field potential.
Although the general trend of our observations is well reproduced
by this model, finer details are not. To take interactions within
the thermal cloud and between the thermal cloud and the condensate
into account, we use a self-consistent Hartree-Fock (HF)
description of the mixed cloud
\cite{goldman1981a,huse1982a,oliva1989a,bagnato1988a,shi1997a,shi1997b,giorgini1997b},
which yields good agreement with the data. We are able to confirm
experimentally its validity over a wide range of temperatures and
atom numbers, and to show that despite the diluteness of the
thermal cloud, its mean-field energy affects both the condensed
and non-condensed atoms. Such systematic measurements of
the mutual interplay between the condensed and non-condensed
components have not been reported before, although evidence for 
repulsion of the thermal atoms by the condensate has been provided in \cite{busch2000a}, 
through a careful study of the interface region btween the two components.

Our measurements rely on the standard time-of flight technique to
observe the mixed cloud. A limitation arises at low temperatures,
where the condensate appears broader than the thermal cloud in
time of flight. This limits our measurements with this technique
to $T/T_{\rm c}^0
> 0.3$ (condensed fraction $< 80$ \%). To overcome this
limitation, we use coherent Bragg scattering to spatially separate
the coherent and incoherent components of the cloud in time of
flight. We apply it for low temperature thermometry, and measure
in this way condensed fraction larger than 90 \% and temperatures
below 100 nK ($T/T_{\rm c}^0 \approx 0.2$).

The paper is organized as follows. Section \ref{methods} first
describes the experimental apparatus used for condensate
production. Then we expose the widely used procedure of
time-of-flight imaging, and introduce the improved method based on
Bragg diffraction to filter the condensate out of thermal cloud.
The key results of this paper are presented in sections
\ref{sectionfc} and \ref{sectionErel}. In section \ref{sectionfc},
we present a measurement of the temperature dependance of the
condensed fraction. In section \ref{sectionErel}, the Bragg
filtering scheme is applied to a measurement of the expansion
energy of the thermal component. Both measurements clearly confirm
the HF predictions in the whole temperature range. Further
evidence for mutual interaction is provided in section
\ref{sectionexp}, where we analyze the behavior of the mixed cloud
in time of flight and measure a compression of the condensate
axial length by the thermal cloud. For completeness, we review
briefly in appendix \ref{theory} the three well-understood models
to which we compare the data, namely the ideal gas model, the
semi-ideal model and the self-consistent HF model.
\section{Experimental methods}\label{methods}
This section reviews the experimental techniques used in this
work. After a brief description of our experimental apparatus
(\ref{bec_production}), we discuss in section \ref{mesT} condensed
fraction and temperature measurements via time-of-flight
absorption imaging, and point out to the limitations of this
technique for high condensed fractions. In section
\ref{bragg}, we show how they can be overcome using coherent Bragg
scattering \cite{kozuma1999a,stenger1999b}.

\subsection{Condensate production}\label{bec_production}
Our experimental setup employs the standard combination of laser
and evaporative cooling to reach Bose-Einstein condensation in a
sample of $^{87}$Rb atoms, spin-polarized in the $|F=1;m_{\rm
F}=-1\rangle$ hyperfine ground state. A Zeeman-slowed atomic beam
loads a magneto-optical trap in 4 seconds. After spatial
compression (125 ms) and cooling in optical molasses (6 ms), the
laser cooled sample is repumped in the $\mid F=1\rangle$ hyperfine
manifold and loaded into an Ioffe-Pritchard type magnetic trap.
The trap, an iron-core electro-magnet, is a specific feature of
our experiment \cite{desruelle1998a}. The guiding of magnetic flux
lines along the ferromagnetic body of the magnet produces strong
(when compared to macroscopic, coil-based devices) radial
gradients (1.4 kG/cm), while using a modest current of 30 Amps.
This produces a cigar-shaped, very-elongated harmonic potential of the form $V_{\rm
ext}=M\omega_\perp^2(x^2+y^2)/2+M\omega_z^2 z^2/2$. In this work, differently from \cite{richard2003a},
radial and axial trapping frequencies are respectively
$\omega_{\perp}/2\pi=413(5)$\,Hz and $\omega_{z}/2\pi=8.69(2)$\,Hz
at a 10 G bias field.

The combined compression and evaporation sequence typically lasts
for 16 seconds, with the last part of the radio-frequency (rf)
evaporation ramp considerably slowed down (to a ramp speed of
$200$\,kHz/s) and followed by a $1$\,s hold time in the presence
of an rf shield. This is done in order to allow the system to
cross slowly the quantum degeneracy threshold, and to relax
towards equilibrium. In particular, non-equilibrium shape
oscillations that occur in such anisotropic traps upon
condensation \cite{richard2003a,shvarchuck2002a} are strongly
reduced. To ensure the reproducibility of the temperature in the
experiment, the radio-frequency $\nu_{\rm 0}$ that empties the
trap is measured every five experimental cycles, and the rf shield
frequency $\nu_{\rm rf}$ is readjusted in real time to follow slow
drifts of the bottom of the trap. In this way, the ``trap depth''
$\nu_{\rm rf}-\nu_{0}$ is controlled within $\pm 2$\,kHz. Since we
measure $h(\nu_{\rm rf}-\nu_{0})/k_{\rm B} T\approx$ 11 in this
final evaporation stage, we estimate the temperature
reproducibility to be $\pm 10$\,nK.
\subsection{Time of flight imaging}\label{mesT}
Informations on the atomic cloud are obtained through
time-of-flight absorption imaging (see \cite{Ketterle1999a} for a
detailed account of imaging techniques, and \cite{imaging} for
details of the implementation in our experiment). This widely
employed method will be referred to as the ``standard method'' in
the following. Here, two important quantities are measured through
a fit to absorption images, the condensed fraction and the
temperature. To find the condensed fraction, we use the fitting
method described in \cite{Ketterle1999a}, which assumes that the
atomic density can be described by a parabolic TF profile
describing the condensate plus an ideal Bose-Einstein distribution
with zero chemical potential describing the thermal cloud. The
condensed number $N_0$ is taken to be the number of atoms
integrated under the parabolic profile, while the total atom
number $N$ is found by integration over the entire profile.

The temperature is measured through an independent fit to a
thermal Bose-Einstein distribution, restricted to the wings of the
thermal distribution only, in order to minimize the effect of
interactions (see \cite{ensher1996a,mewes1996a} and section
\ref{sectionErel}). One typically assumes that the high-energy
atoms selected by this procedure behave as if the gas were ideal,
and extracts effective temperatures $T_{x}$ and $T_{ z}$ from the
cloud sizes, for instance $k_{\rm B} T_{x} = M \omega_{\perp}^2
R_{\rm th}^2/[1+(\omega_\perp t)^2]$ with $R_{\rm th}$ the radial
size. We observe a systematic variation of the temperature $T_{
z}$ measured along the long axis with the size of the exclusion
region. This effect was more pronounced with increasing condensed
fraction, and disappeared above $T_{\rm c}$. On the contrary, the
radial temperature $T_{x}$ was barely affected by the actual size
of the exclusion region, provided it was chosen larger than the
condensate radius and sufficiently small to conserve a reasonable
signal-to-noise ratio (5\% or less variation). For this reason, we
infer the initial temperature from the radial value $T_{x}$ only
\cite{notetemp}. This thermometry procedure assumes a
nearly-ballistic expansion. Collisional effects in time-of-flight
may invalidate this assumption, and it is important to estimate
their importance to quantify the accuracy of our measurements. We
defer this discussion to sections \ref{sectionErel} and
\ref{sectionexp}.
\subsection{Selective displacement of the condensate using coherent
Bragg scattering}\label{bragg}
The method described in the last subsection rests on the clean
distinction between the condensed and thermal components. However,
for cold samples ($k_{\rm B}T \lesssim \mu$), the condensate
radius after time of flight is larger than the extension of the
thermal cloud. The details of the thermal distribution are
therefore ``buried'' under the condensate density profile, and a
reliable fit is not possible. We have found no difficulties for
$N_0/N \sim 60$\% ($T/T_{\rm c}^0\sim0.5$), while still being able
to extract valuable informations for condensed fractions close to
80 \% ($T\gtrsim 0.3~ T_{C0}$). For higher condensed fractions,
the signal-to-noise on the wings is too low to perform a reliable
fit.

\begin{figure}
\includegraphics[width=5cm]{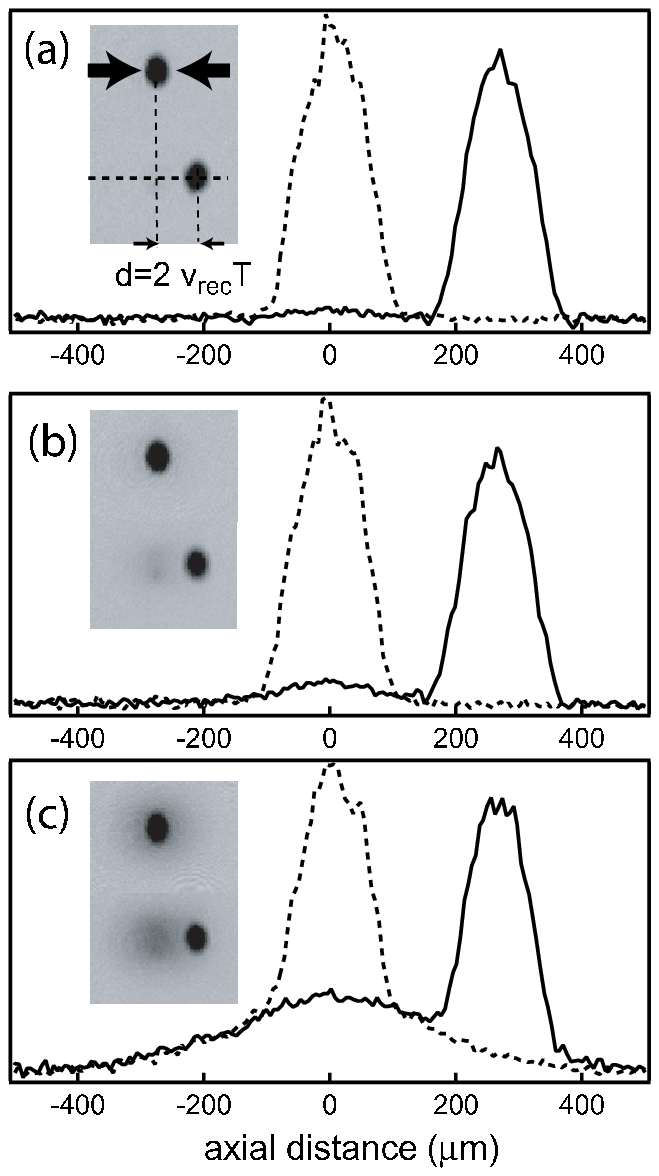}
\caption{Bragg diffraction as a condensate filter. Figures {\bf
(a)}, {\bf (b)} and {\bf (c)}correspond to 92, 85 and 30 \%
condensed fraction, respectively ($T/T_{\rm c}^0 \approx$0.2, 0.35
and 0.8). Two-dimensional absorption images are shown, with a cut
along the direction of the trap weakest axis (``axial cut''). The
top images show regular absorption images for a 24.27 ms time of
flight (dashed lines in the cut). The bottom images, also shown as
the solid line in the axial cut, corresponding to the same
temperatures and atom numbers within experimental reproducibility,
have been taken after applying a moving optical lattice tuned to
realize a $\pi$ Bragg pulse, transferring two photon recoils to
almost all condensed atoms (velocity 1.1 cm/s). The distance
travelled by the condensate is verified to be 250 $\mu$m,
corresponding to a free flight of 22.27 ms after the Bragg pulse.
As can be seen, the thermal cloud is barely affected, due to its
much larger extent in momentum space.} \label{bragg_pics}
\end{figure}

In this section, we describe an improvement of the time-of-flight
technique that exploits the dramatic difference between the two
components in momentum space
\cite{dalfovo1999a,kozuma1999a,stenger1999b}. Our primary
motivation for introducing this method is to improve the
resolution at low temperatures, a key advantage in the condensed
fraction measurements presented below. The principle of the
technique is to transfer the whole condensate to a center-of-mass
momentum state with two photon recoil velocities (approximately
$1.1$ cm/s in our case), while leaving the thermal atoms
essentially unaffected. As shown in Fig. \ref{bragg_pics}, after a
subsequent time of flight (typically 20 ms), the condensate has
moved away from the center of the thermal cloud, allowing us to
perform separate fits and revealing previously hidden information
about the thermal component. This technique is related to the
filtering scheme used in \cite{ferlaino2002a} to study collective
modes in the presence of a weak optical lattice at finite
temperatures.

Momentum transfer is achieved by exposing the sample to two
counter-propagating laser beams detuned with respect to each
other. The moving lattice formed this way can Bragg diffract the
atoms, promoting them to a two-recoil momentum state. In
principle, Bragg scattering is sensitive to the atomic momentum
distribution \cite{kozuma1999a,stenger1999b}. However, if the
spectral width of the Bragg pulse, Fourier-limited by the pulse
duration $T$ to $\Delta \nu_{\rm F} \sim 1/T$, is much larger than
the condensate Doppler width $\Delta \nu_0 \sim 2 v_{\rm R}/L$,
the momentum sensitivity of Bragg diffraction is lost and almost
complete transfer of the condensate ($\pi$ pulse) is achieved. On
the other hand, if the thermal component spectral width, $\Delta
\nu_{\rm th} \sim 2 v_{\rm R}/\lambda_{\rm T}$ ($\lambda_{\rm
T}=\sqrt{2 \pi\hbar^2/m k_{\rm B} T}$ is the thermal De Broglie
wavelength), is much larger than $\Delta \nu_{\rm F}$, Bragg
diffraction is then a momentum-selective process, and most of the
thermal atoms, being off-resonant with the Bragg beams, stay at
rest in the laboratory frame. This filtering technique thus makes
possible independent manipulation of the condensed and
non-condensed atoms.

The experimental setup that we use to generate the Bragg beams is
identical to the one used in \cite{richard2003a}. The $\pi$ Bragg
pulse is applied after sudden switch-off of the trapping potential
and a 2 ms time of flight, to avoid s-wave collisions between
atoms with different momenta
\cite{chikkatur2000a}. The Bragg beams are counter-propagating and
parallel to the elongated axis of the trap. Since the condensate
mean-field energy is then almost entirely released in the
transverse directions, perpendicular to the diffraction axis, the
ratio between the condensate and thermal cloud spectral widths is
\begin{equation} \label{deltapratio}
\left(\frac{\Delta \nu_0}{\Delta \nu_{\rm th}}\right)_{\rm t.o.f.}
\sim \left(\frac{\omega_z}{\omega_\perp}\right)^2\sqrt{
\frac{\mu}{k_{\rm B} T}},
\end{equation}
and remains smaller than one even at very low temperatures. Note
that this would not have been the case for Bragg beams parallel to
a radial axis, where the momentum width is larger by a factor
$(\omega_\perp/\omega_z)^2$ due to the the released mean field
energy.

To find the condensate fraction, we simply count the number of
diffracted atoms and identify it with the condensed number. To
obtain a meaningful measurement of the condensed fraction, two
conditions have to be met. First, almost all condensate atoms
should be diffracted. Transfer efficiencies as high as 97 \% were
observed for the coldest clouds we have produced, indicating that
only a few percent of the condensate population remains at rest;
this is comparable to the $\pm 2$ \% root-mean-square fluctuations
of the diffraction efficiency that we have measured
experimentally. Second, the diffracted number of thermal atoms
should be small to obtain a sensible measurement of the condensed
number. We estimate the diffracted {\it fraction} of thermal atoms
$f_{\rm th}$ in the impulse approximation discussed in
\cite{zambelli2000a}. In this approximation, well justified for
the low densities considered here, the number of atoms diffracted
from the thermal cloud is taken to be proportional to the thermal
part of the momentum distribution, integrated over the directions
perpendicular to the axis of the Bragg beams. Treating the thermal
cloud as an ideal Bose gas in a first approximation, we find for a resonant $\pi$ pulse
\begin{equation}
f_{\rm th}=\frac{\pi^2}{4\Delta \nu_{\rm th} T} \frac{\left[
g_{5/2}(1) - g_{5/2}\left\{\exp \left[-16(\nu_{\rm
R}/\Delta\nu_{\rm th})^2 \right] \right\} \right]}{g_{3/2}(1)}.
\end{equation}
Here and below, the Bose functions are defined by
$g_{\alpha}(u)=\sum_{j \geq 1}u^{j}/j^{\alpha}$. The second term
in brackets arises from off-resonant excitations, which are
negligible for the condensate but sizeable for the thermal
component. The quantity $f_{\rm th}$ varies from 6 \% at 400 nK
($T \sim 0.8~T_{\rm c}^0$, or $N_0/N \sim 20$ \%) to 20 \% at 150
nK ($T \sim 0.2~T_{\rm c}^0$, or $N_0/N \sim 90$ \%). The number
of thermal atoms in the diffracted peak is given by
$(1-N_0/N)f_{\rm th}$. All in all, we estimate from this simplified calculation
that the Bragg filter is a useful technique in a window  $5\% < N_0/N < 95 \%$,
the upper bound being set by the first condition (complete
condensate transfer) and the lower by the second (number of
condensed atoms larger than number of thermal atoms in the
diffracted peak). Outside of this window, it becomes suspect to
identify unambiguously the diffracted order with the condensate.
Despite of this, the technique represents a significant
improvement over the standard method, as we will see in the next
sections.
\section{Condensed fraction of the interacting gas versus temperature}\label{sectionfc}
With the tools of section \ref{methods} in hand, we can
investigate the behavior of the condensed fraction as a function
of temperature. This quantity is of primary importance: the
possibility to measure it directly in trapped gases is in stark
contrast with the situation in traditional, bulk superfluids where
such measurements are intrisically difficult \cite{sokol1993a}. We
have done this measurement in two steps, first by using the
standard method and second by taking advantage of the enhanced
resolution of the Bragg filtering scheme.

\subsection{Standard time-of flight
measurements}

\begin{figure}[!t]
\includegraphics[width=15cm]{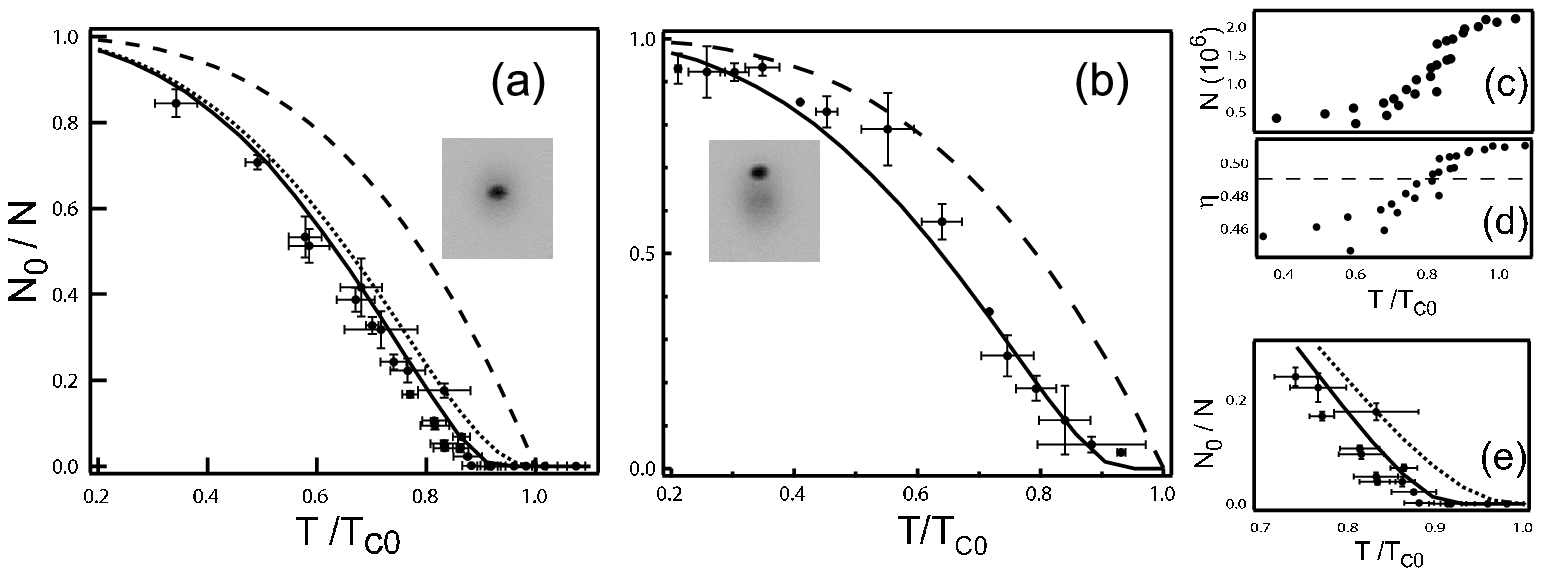}
\caption{Condensed fraction as a function of reduced temperature.
{\bf(a),(b)} Experimental data after averaging are shown as filled
circles, with statistical error bars. Data in {\bf (a)} were taken
using the standard time-of-flight technique, while data in {\bf
(b)} were measured with the Bragg filtering scheme and extend to
lower temperature, down to $T \approx 0.2~T_{\rm C0}$ whereas the
standard method is limited to $T \approx 0.4~T_{\rm C0}$. Lines
show theoretical expectations according to an ideal gas
calculation, including finite-size effects (long dashed), a
``semi-ideal'' model that neglects interactions within the thermal
cloud (short dashed), and a self-consistent HF calculation
(solid). Due to a different average atom number, the parameter
$\overline{\eta}$ that controls the importance of interactions is
different in each case, $\overline{\eta}=0.49$ and
$\overline{\eta}=0.47$ for {\bf (a)} and for {\bf (b)},
respectively. The variation in total number due to evaporative
cooling across the data set shown in {\bf (a)} is shown in {\bf
(c)}, and the corresponding variations of $\eta$ in {\bf (d)} 
(Note the vertical scale, extending over no more than 5 \% of the average value). 
The average $\overline{\eta}=0.49$ is shown by the dashed line. In
{\bf (e)}, we show an enlargement of {\bf (a)} around the critical
temperature, to highlight the importance of making the full HF
calculation to reproduce the trend seen in the data.}
\label{cond_frac}
\end{figure}

Using the standard analysis technique ({\it i.e.} no Bragg
filtering), we investigate in this section a temperature interval
ranging from $T \approx 1.1~ T_{\rm c}^0$ down to $T \approx 0.3
~T_{\rm c}^0$, with approximately $2\times10^6$ atoms at the
transition. Fig.~\ref{cond_frac} shows the measured condensed
fraction as a function of $T/T_{\rm c}^0$. Each point results from
an averaging over several (typically five) realizations under
identical conditions. The expectation for an ideal gas, including
finite-size effects \cite{dalfovo1999a}, lies distinctly above our
experimental data. The difference can be attributed to
interactions, as shown by the far better agreement with the
self-consistent HF calculation (solid line). The observed
reduction of the condensed fraction contrasts with the homogeneous
case \cite{giorgini1996a}, where the condensed and normal
component overlap everywhere, and where it is energetically
favorable to {\it increase} the condensed fraction to diminish the
exchange interaction energy among excited states.

An interesting property, first pointed out in
\cite{giorgini1997a}, is the scaling behavior of all
thermodynamical quantities, which depend only on the reduced
temperature $T/T_{\rm c}^0$ and on the parameter $\eta$ which
controls the magnitude of two-body repulsion \cite{semicl},
\begin{equation}
\eta=\frac{\mu_{\rm TF}[N_0=N]}{k_{\rm B}T_{\rm c}^0} \approx 1.57
\left(\frac{a}{\overline{\sigma}}\right)^{2/5} N^{1/15}.
\end{equation}
Alternatively, one can express $\eta$ as the ratio between the two
characteristic lengths in the uniform problem, the scattering
length $a$ and the De Broglie wavelength $\lambda_0$ for $T=T_{\rm
c}^0$, as $\eta \approx 1.07 (a/\lambda_0)^{2/5}$. The power $2/5$
reflects the presence of the trapping potential. A typical value
in our experiment is $\eta=0.5$, while the experiment of
\cite{ensher1996a} (for instance) corresponds to
$\eta\sim0.3-0.4$. Thus, the effect of two-body interactions is
stronger in the work reported here, which explains to some extent
the clarity with which we observe deviations from ideal gas
behavior.

In the experiment, because the total number $N$ drops with $T$ due
to evaporation and losses, $\eta$ slightly decreases across the
temperature range, from $0.51$ above $T_{\rm c}$ to $0.47$ at low
temperature (see Fig. \ref{cond_frac}c and d). To compare with
theory, we use the average number of atoms across the data set,
$\overline{N}\approx1.2\times10^6$, and the corresponding
$\overline{\eta}=0.49$. This does not lead to a discernible
variation of the HF prediction at the scale of the graph, because
of the very weak $N^{1/15}$ dependence of $\eta$. This behavior
emphasizes the scaling behavior exhibited by trapped Bose gases
with large atom numbers \cite{giorgini1997a}.

Finally, we note that the data in the vicinity of $T_{\rm c}$,
shown in Fig.~\ref{cond_frac}e, emphasize the necessity of the
full HF treatment of the thermal component to understand
quantitatively the thermodynamic properties. As a matter of fact,
the semi-ideal model (dashed line) predicts a condensed fraction
systematically higher than the one we observe. This is clear
evidence for mutual interaction between the condensed and
non-condensed components. We will return later to this point,
which is an important conclusion to be drawn from this work.

\subsection{Enhanced resolution of low thermal fraction}

As discussed earlier, at low temperatures, $T\lesssim0.4 T_{C}$,
the standard procedure is unable to extract faithfully the
properties of the thermal cloud. Thanks to the Bragg filtering
technique introduced in section \ref{bragg}, this difficulty can
be overcome and very small thermal fraction can be detected. The
condensed fraction measured this way is plotted in Fig.
\ref{cond_frac}b, along with the $\overline{\eta}=0.47$ curve that
corresponds the value $\overline{N}\approx8\times10^5$ for this
set of measurements. Again, we find good agreement with the
prediction of the HF model within our uncertainty, even at very
low temperatures. One could wonder whether the contribution from
collective excitations (quantum and low-energy thermal depletion)
could be measured by this technique (the analogous of the phonon
regime in superfluid $^4$He). Unfortunately, according to the
estimations of \cite{giorgini1997b}, they are always small when
compared to the contribution of single-particle excitations in the
temperature range we explore, and compared to the estimated
sensitivity of the Bragg technique (section \ref{bragg}). Only by
increasing the diluteness parameter $\sqrt{n_0({\bf 0}) a^3}$
significantly could this regime become observable experimentally with
the techniques described here.

\section{Measurement of the overall and thermal expansion energies}\label{sectionErel}
Another quantity that can be measured from time-of-flight
expansion is the release energy
\cite{ensher1996a,mewes1996a,dalfovo1999a}, the sum of the kinetic
and interaction energy released at the trap cut-off and available
for the expansion of the whole cloud. In an anisotropic trap such
as ours ($\omega_\perp/\omega_z\approx51$), almost all the
interaction energy converts into radial expansion velocity. The
radial expansion of the cloud for $t\gg\omega_\perp^{-1}$,
observed in the $y$ direction, proceeds at an overall speed
$v_{y}$, fixed by the expansion energy
\begin{eqnarray}
E_y=\frac{1}{2}M v_{y}^2 = \frac{1}{3}E_{\rm
kin}+\frac{1}{2}E_{\rm int},
\end{eqnarray}
which, if scaled by the characteristic $N k_{\rm B}T_{\rm c}^0$,
is a universal function of $\eta$ and $T/T_{\rm c}^0$. We discuss
in Appendix \ref{theory} how to calculate the kinetic and
interaction energies in the HF approximation.

\begin{figure}
\includegraphics[width=14cm]{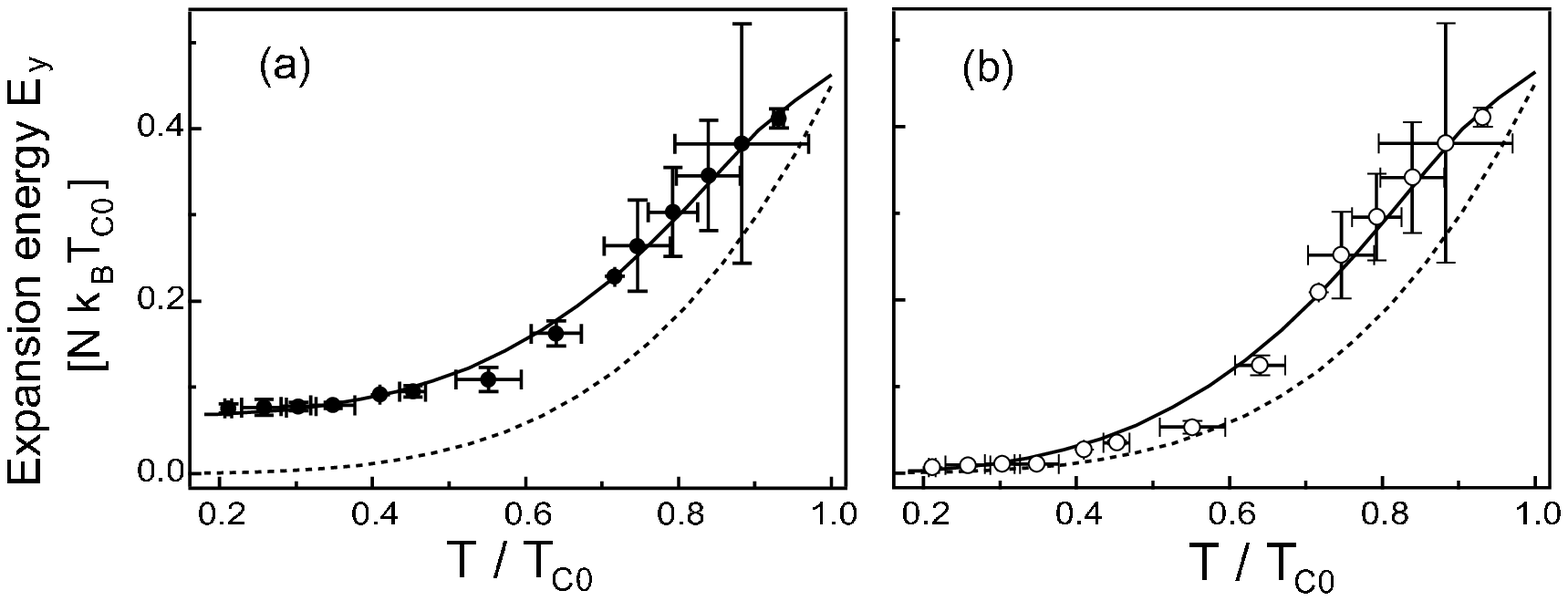}
\caption{Expansion energy in the radial $x$ direction as a
function of temperature. Filled circles in {\bf (a)} correspond to
the whole cloud, and open circles in {\bf (b)} to the thermal
component only. The data are taken from the same set as in Fig.
\ref{cond_frac}b. The solid lines on the graphs show the same
quantities predicted by the self-consistent HF model, with
$\eta=0.47$. The dotted line is the expansion energy (kinetic
only) of an ideal cloud.} \label{Erel}
\end{figure}

Experimentally, one measures $E_{y}$ directly from the root mean
square cloud radius, according to $\langle y^2 \rangle = v_{y}t$
\cite{ensher1996a,mewes1996a}, without resorting to a detailed
fitting model. This expression assumes negligible relaxation
between the axial and radial degrees of freedom. We have plotted
the measured value of $E_y$ in Fig. \ref{Erel}a. for same data as those 
shown in Fig. \ref{cond_frac}b, where the Bragg filter has been
used ($\eta\approx0.47$ for these data). As expected, the kinetic
energy of the thermal cloud dominates close to $T_{\rm c}$, with a
small contribution of the mean-field energy, whereas the
interaction energy of the condensate $E_0$ is the most important
term at low temperature. The HF curve connects these two limiting
cases, and reproduces well our observations.

A key advantage in our situation is the ability to analyze
separately the condensate and the thermal cloud thanks to the
Bragg filtering scheme, and therefore to measure the release
energy of the thermal cloud {\it alone}. To avoid the condensate,
we estimate the thermal cloud rms radius from a fit to a radial
cut to the image (see inset in Fig. \ref{Erel}b). The release energy
of the thermal cloud measured this way is shown in Fig.
\ref{Erel}b, together with the calculated value,
\begin{eqnarray}
E_y^{\rm (th)}= \frac{1}{3}E_{\rm kin}+\frac{k_{\rm B} T}{2}
(\zeta_0+\zeta_{\rm th}).
\end{eqnarray}
The quantities $\zeta_0$, $\zeta_{\rm th}$ represent the scaled
mean-field energy corresponding to the repulsion felt by a
thermally excited atom due the condensate and to the remaining
thermal atoms, respectively. They are defined more precisely in
Appendix \ref{theory}. Although the difference with the
non-interacting curve is less pronounced at very low temperatures,
where the thermal energy is very low, close to $T_{\rm c}$, these
results emphasize the important role of interactions and the good
agreement with HF theory once again.

\section{Further evidence for mutual interaction between the condensed and thermal
components}\label{sectionexp}

\begin{figure}
\includegraphics[width=6cm]{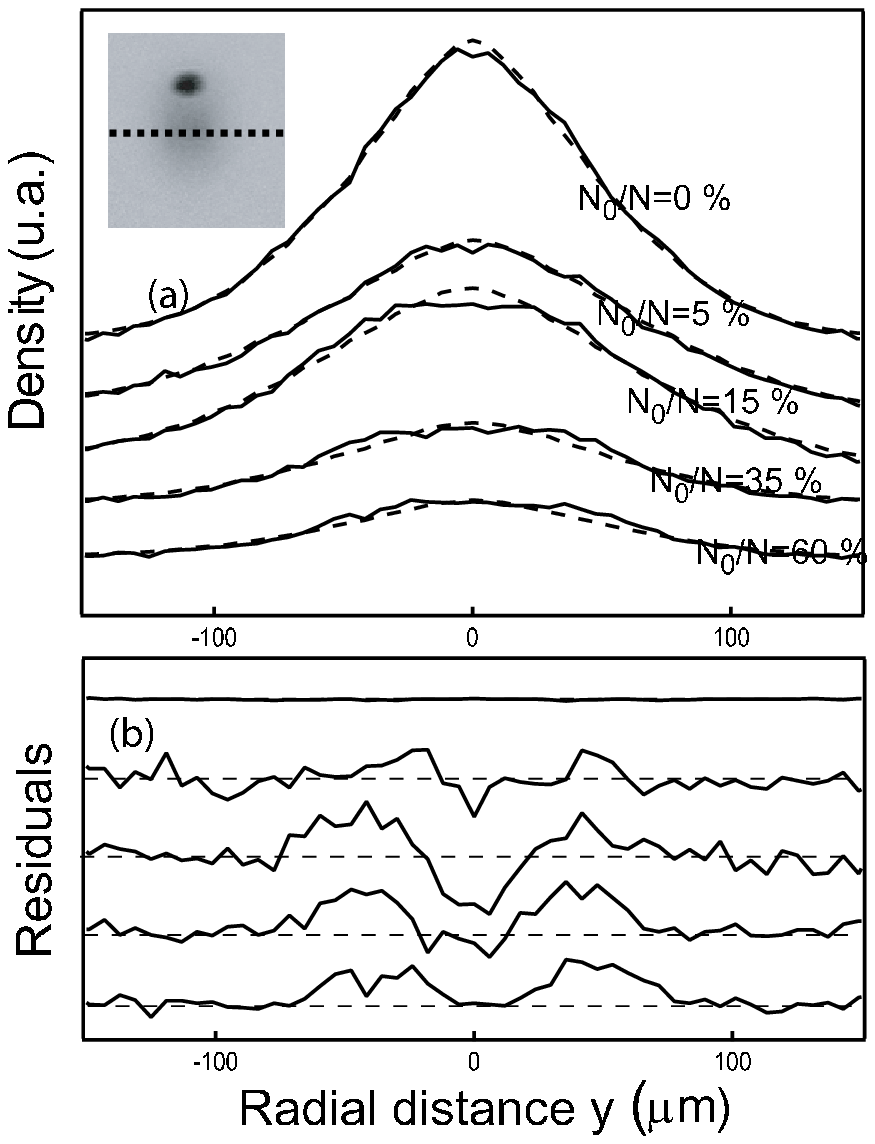}
\caption{Radial density profile of the thermal cloud. The dashed
line in the inset shows the axis along which the profile are
taken. {\bf (a)} Radial cut of the column density profile of the
thermal cloud after 24.3 ms time of flight (solid) and the best
fit to an ideal Bose-Einstein ditribution (dashed). The Bragg
filter has been employed to separate the thermal and condensed
components. The condensed fraction is indicated in each case. {\bf
(b)} Residual of the fit for each case in {\bf (a)}.}
\label{profils}
\end{figure}

From images taken employing the Bragg filtering scheme, it is also
possible to examine a radial cut to the profile of the thermal
cloud, as shown in Fig. \ref{profils}a. As the initial kinetic
energy of the thermal cloud is typically much larger than its
mean-field energy, one expects that the density distribution after
time of flight reflects at least approximately the initial
momentum distribution, which does not display the ``hole'' present
in the density distribution in the trap (see Appendix \ref{theory}
for further discussion). This is indeed the case: the measured
profiles show a monotonic behavior near the center of the cloud
\cite{visibhole}. Nevertheless, the density distribution we
observe is somewhat flatter in the central region than the ideal
gas distribution used in the analysis, as shown by the residuals
of a fit to the radial profile (Fig. \ref{profils}b). Above
$T_{\rm c}$, the effect disappears, which indicates that a flatter
profile is not simply an artifact of our measurement method.
Furthermore, such a behavior is to be expected if the condensate
mean field repels the thermal cloud in the early stage of the
expansion, since interactions tend in general to make the density
profile more uniform.

\subsection{Evidence for non-ballistic expansion}\label{expansion}

\begin{figure}[ht!]
\includegraphics[width=16cm]{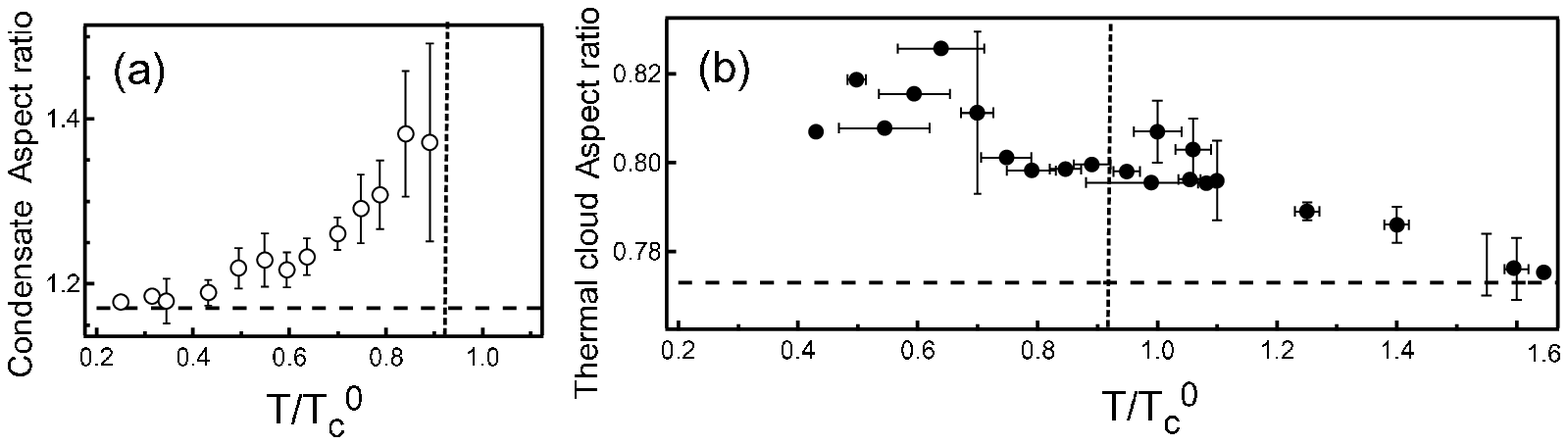}
\caption{Deviation from ballistic expansion. The aspect ratio of
the condensed {\bf(a)} and thermal {\bf(b)} components of mixed
clouds, after 22.3 ms of free expansion, are plotted as a function
of the reduced temperature. The aspect ratio for non-condensed
clouds, analyzed in more details in \cite{gerbier2003c}, are also
shown for comparison. The horizontal dashed lines indicate the
aspect ratio of a TF condensate and an ideal thermal gas,
respectively, and the vertical dotted lines show the critical
temperature including mean-field and finite-size effects.}
\label{arT}
\end{figure}

As stated in section \ref{methods}, this repulsion effect is not taken into account
in our fitting procedure, which assumes an ideal Bose-distribution
to fit the profile and ballistic expansion to deduce temperature
from the cloud sizes. To investigate further the validity of the
analysis, we begin by plotting in Fig. \ref{arT}a and \ref{arT}b
the aspect ratio of both components, as a function of the reduced
temperature. One sees from these graphs that the simple model of a
TF condensate on top of an ideal thermal background is not
sufficient to account for the data. Indeed, for our trapping
frequencies and for the time of flight $t=22.3$ ms used in these
measurements, one would expect from this model an aspect ratio of
1.17 for the condensate and $0.77$ for the thermal cloud. Both
deviate from these values and vary with temperature, indicating
that the expansion dynamics is more complex than assumed by the
analysis model. Note that although this model fails to describe
fully the expansion dynamics, the observed deviation from
ballistic expansion remains small.

As already pointed out, the obvious weakness of the analysis model
is the neglect of collisional effects. We recall that the
expansion model is basically motivated by the absence of a more
elaborate theory to which we could compare our observations. It is
however of interest to quantify the error level on temperature
measurements, which we will do in the remainder of this section at
the gross estimate level. Interaction-driven forces that affect
the expansion can be divided into two distinct classes
\cite{pedri2003a}: hydrodynamic forces on one hand, predominant
above $T_{\rm c}$ in \cite{gerbier2003c}, and mean-field repulsion
on the other, which play a minor role above $T_{\rm c}$, but
becomes increasingly important with decreasing $T$ due to the
presence of the condensate \cite{dyndep}. Both effects are roughly
speaking comparable in magnitude, and from our observations above
$T_{\rm c}$ and HF calculations (see section \ref{theory}), we
estimate an upper bound on the systematic error on temperature
measurement, from $10$ \% close to $T_{\rm c}$ (mostly due to
hydrodynamic behavior) to $20$ \% well below (mostly due to
repulsion by the condensate), compatible with the observed
deviation from the ideal gas expansion.

This is an upper bound because the temperature is found through a
fit to the wings of the distribution the only, which is expected
to reduce effect of mean-field repulsion. The close agreement we find between the
calculated release energy and the measured one in Fig.
\ref{Erel}b, where we recall that the
radial size used to deduce the expansion energy was found through
a fit to the full profile, seems to indicate that
fitting to the wings avoids counting most of the repulsion energy
in the determination of the temperature, as expected. However,
this need not be true for the hydrodynamic relaxation, which
affects directly the momentum distribution. Just as above $T_{\rm C }$,
hydrodynamic forces are thus likely to be the dominant source of
systematic error in most of the temperature range in
elongated traps such as ours.

\subsection{Compression of the condensate by the thermal cloud}
We conclude this section by examining the axial length of the
condensate, which is reasonably immune to the effects discussed
above because the axial expansion of the condensate is very slow:
the measured length thus stays close to the in-trap length. From
condensed fraction measurements, it is clear that the semi-ideal
model is not sufficient to reproduce our results, meaning that the
mutual interaction between condensed and non-condensed atoms is
observable. This also seen from the length of the condensate
inferred from the two-component fit, which is reduced when
compared to the TF length calculated with the number of condensate
atoms we measure. A quantitative comparison can be made only by
taking into account the slow axial expansion. In the absence of a
complete theory, we assume that the axial length is rescaled from
the equilibrium length by the same factor as a condensate in the
TF regime, $b_{z}\approx 1 + \pi \omega_z^2
t/2\omega_\perp\approx$ 1.04 for our parameters
\cite{castin1996a,Kagan1997b}. With this assumption, we find our
data to be in reasonable agreement with the HF length, whereas the
TF prediction is found systematically too high (see Fig.
\ref{becL}). This reduction can be attributed to the thermal cloud
compression discussed in appendix \ref{theory}: at equilibrium,
the shell of thermal atoms surrounding the condensate exerts a
force towards the trap center, reducing its extension when
compared to a ``free'', TF condensate.

\section{Conclusion}
In this paper, we have investigated experimentally the
thermodynamics of a trapped, interacting Bose gas over a wide
range of temperatures, from $T_{\rm c}$ down to  $0.2~T_{\rm c}$.
We have used the standard time-of-flight analysis, complemented by
the use of coherent Bragg scattering to filter the condensate out
of the thermal cloud. The latter technique allows us to reach
lower temperatures and higher condensed fractions than those
accessible by the usual method. We have investigated primarily two
quantities, the condensed fraction and the (radial) release
energy. The data display without ambiguity an interacting gas
behavior, and are in agreement at a few percent level with a
Hartree-Fock description of the mixed cloud. This highlights the
pertinence of this simple description of the interacting, trapped
Bose gas at finite temperature. Moreover, this gives evidence for
mutual interaction between the condensate and the thermal cloud
close to $T_{\rm c}$. Although these effects are small, they are
measurable and should be taken into account in precise comparisons
to the theory of finite temperature, Bose-Einstein condensed
gases.

A more thorough quantitative test of theory for ground state
occupation and similar thermodynamic properties is however
hindered by the lack of theory to understand the expansion of a
mixed cloud. For instance, the data in Fig.~\ref{cond_frac}a lie
systematically slightly below the theory, which may be due to an
underestimation of the condensed fraction or the temperature: as
described above, we use an ideal Bose distribution to fit the
thermal component. This assumption contradicts the principal
conclusion of this paper, that ideal gas approximations are far
from sufficient at the desired level of accuracy. Although the
experiments described in this paper suggest an accuracy on temperature of order 10 \% at
least, we have also shown that the density profile deviates from an
ideal gas near the center of the cloud, and stressed that
hydrodynamic and mean-field effects in the expansion are not
properly accounted for. Systematic errors may thus still be
present, and a better accuracy is not guaranteed with the methods
used in this paper. Numerical work, for instance along the lines
of \cite{zaremba1999a}, may help study the expansion of mixed
clouds and improve the standard analysis procedure.
\begin{acknowledgments}
We would like to thank D. Gu{\'e}ry-Odelin, S. Giorgini, J.
Dalibard, G. V. Shlyapnikov, J. Retter and D. Boiron for useful
discussions and comments on the manuscript, as well as D.
Cl\'ement for his work on a related project. JHT acknowledges
support from a Ch{\^a}teaubriand fellowship, from CNRS and from NSERC, and MH
from IXSEA. This work was supported by D\'el\'egation G\'en\'erale
de l'Armement, the European Union (Cold Quantum Gas network),
INTAS (contract 211-855) and XCORE.
\end{acknowledgments}
\appendix
\section{Simple models of a trapped Bose-Einstein condensed gas at finite temperatures}\label{theory}
In this Appendix, we briefly summarize the three different models
to which we compare our observations. We use as a temperature
scale the critical temperature of an ideal gas in the
thermodynamic limit\cite{dalfovo1999a}, $k_{\rm B}T_{\rm
c}^0=\hbar\overline{\omega}[N/\zeta(3)]^{1/3}$, where $\zeta$ is
the Riemann function, and
$\bar{\omega}=\omega_{\perp}^{2/3}\omega_{ z}^{1/3}$ is the
geometrical mean of the trapping frequencies. We make three key
assumptions to simplify the theoretical description. For the
condensate, we suppose that the condition $N_0 a /
\overline{\sigma} \gg 1$ holds, where
$\overline{\sigma}=\sqrt{\hbar/M\bar{\omega}}$ is the mean ground
state width and $a$ the scattering length. This ensures that the
condensate is in the TF regime. At very low temperatures, where
the condensed fraction is almost unity, this gives the condensate
density $n_0=n_{\rm TF}$ as
\begin{eqnarray}n_{\rm TF}({\bf{r}}) = [\mu_{\rm TF}
- V_{\rm ext}({\bf{r}})]/U,
\end{eqnarray}
where $U=4\pi\hbar^2a/M$ is the mean-field coupling constant. The
TF chemical potential is $\mu_{\rm
TF}=(\hbar\bar{\omega}/2)\times(15 N_0 a /
\overline{\sigma})^{2/5}$. For the thermal cloud, we assume
$k_{\rm B} T \gg \hbar \omega_\perp$ for the semi-classical
approximation to hold, and use a mean-field description which
treats thermal atoms as independent particles evolving in a
self-consistent static potential $V_{\rm eff}( {\bf{r}})$ (see below). 
The distribution function in phase space for
thermal particles then reads
\begin{equation}\label{boseeinstein}
f({\bf{r}},{\bf{p}})=\frac{1}{e^{\beta[\mathcal{H}({\bf{r}},{\bf{p}})-\mu]}-1},
\end{equation}
with the semi-classical hamiltonian
$\mathcal{H}({\bf{r}},{\bf{p}})={\bf p}^2/2M+V_{\rm eff}({\bf
r})$, the chemical potential $\mu$, and $\beta=1/k_{\rm B}T$. The
particle density distribution for the thermal component, $n_{\rm
th}$, is found by integegration over momenta and reads
\begin{eqnarray}\label{nth}
n_{\rm th}({\bf{r}}) =\frac{1}{\lambda_{\rm
T}^3}~g_{3/2}\left\{\exp \left[\beta \left(\mu- V_{\rm
eff}({\bf{r}}) \right)\right] \right\}.
\end{eqnarray}
Each model detailed below is thus specified by the precise form of
$V_{\rm eff}$: in the ideal gas model, $V_{\rm eff}$ reduces to
the trapping potential, in the semi-ideal model, it includes the
mean-field of the condensate only, and in the HF model, it also
takes the mean field of the thermal atoms into account.

\subsection{Ideal gas model}

The simplest approximation neglects all interactive contributions
to the effective potential, which reduces to the trapping field,
\begin{equation}
V_{\rm eff}-\mu=V_{\rm ext}-\mu.
\end{equation}
The thermodynamic quantities follow the ideal-gas laws (see for
instance \cite{dalfovo1999a}), and impose in particular $\mu=0$ below $T_{\rm c}$ 
and $\frac{N_0}{N}=1-\left(\frac{T}{T_{\rm c}^0}\right)^3$ for $T \leq T_{\rm
c}^0$. This ideal gas description is thus incompatible with the existence of a TF
condensate, that imply $\mu=\mu_{\rm TF}>0$, and it should be
considered as a reasonable approximation only for $k_{\rm B} T \gg
\mu$.

\subsection{Semi-ideal model}

The repulsion of the thermal cloud by condensed atoms is taken
into account in the so-called ``semi-ideal'' model
\cite{minguzzi1997a,naraschewski1998a}, which considers a TF
condensate containing $N_0<N$ atoms plus a quantum-saturated,
ideal thermal gas moving in the combined trapping plus condensate
mean field potential,
\begin{eqnarray}
\nonumber V_{\rm eff}({\bf r})-\mu& = & V_{\rm ext}({\bf r})+2 U
n_{\rm TF}({\bf r})-\mu_{\rm TF}.
\end{eqnarray}
The factor of 2 accounts for exchange collisions between atoms in
different quantum states \cite{dalfovo1999a}. Note that the
condensed atom number $N_0$ should be found self-consistently for
a given T, under the constraint that the total atom number is
fixed.

The semi-ideal model correctly predicts the most important feature in the static density profile
of trapped interacting Bose gases. As soon as the condensed
fraction is larger than a few percent, the density of the
condensate greatly exceeds the density of the thermal component.
Therefore, the condensate mean field is stronger and repels the
thermal cloud from the center of the trap, digging a hole in the
thermal density distribution.

\subsection{Self-consistent Hartree-Fock model}

Although qualitatively correct and appealing because of its
simplicity, the semi-ideal model is not sufficient to describe
precisely our experiments. Interactions among thermal atoms and
the backaction of the thermal cloud on the condensate have to be
taken into account, which we do here in the HF approximation
\cite{goldman1981a,huse1982a,oliva1989a,bagnato1988a,shi1997a,shi1997b,giorgini1997b},
corresponding to a self-consistent potential
\begin{equation}\label{Veff}
V_{\rm eff}( {\bf{r}})-\mu=V_{\rm ext }({\bf{r}})+2 U n_0
({\bf{r}}) +2 U n_{\rm th}({\bf{r}})-\mu.
\end{equation}
The equilibrium condensate density is no longer simply given by
the TF profile, but depends also on the thermal density through

\begin{equation}\label{gpT}
n_0({\bf{r}})  =\frac{\mu - V_{\rm ext}({\bf{r}}) - 2 U n_{\rm
th}({\bf{r}})}{U}.
\end{equation}
A self-consistent numerical solution with a fixed atom number N
fixes the finite T chemical potential, $\mu=U n_0({\bf 0})+2 U
n_{\rm th}({\bf 0})$, from which all other quantities can be
determined. For instance, the interaction energy, which contains
mixed terms describing the mutual influence of the condensate and
the thermal cloud, can be written as $E_{\rm int}=E_0 + k_{\rm B}
T (2\zeta_0+\zeta_{\rm th})$, where
\begin{eqnarray}
\label{E0} E_{0}&=&\frac{U}{2}\int d^{(3)}{\bf r}~n_0({\bf r})^2,\\
\label{zeta0} \zeta_{\rm 0}&=&\frac{U}{k_{\rm B}T} \int
d^{(3)}{\bf r}~n_0({\bf r})n_{\rm th}({\bf r}),\\
\label{zetath} \zeta_{\rm th}&=&\frac{U}{k_{\rm B}T} \int
d^{(3)}{\bf r}~\left(n_{\rm th}({\bf r})\right)^2.
\end{eqnarray}
The kinetic energy is entirely due to the thermal cloud in the TF
approximation, and reads
\begin{eqnarray}
E_{\rm kin}=\int \frac{1}{h^3}d^{(3)}{\bf r} d^{(3)}{\bf
p}~\frac{{\bf p}^2}{2M}~f({\bf r},{\bf p}).
\end{eqnarray}

Numerical solution of the HF model displays two additional
features compared to the semi-ideal model. First, mean-field
interactions lower the critical temperature for Bose-Einstein
condensation \cite{giorgini1996a,gerbier2003c}. Second, there is a
back-action of the thermal cloud on the condensate: the mean-field
exerted by the shell of thermal atoms surrounding the condensate
acts in return to compress it, increasing its density and reducing
its axial length $L_0$ according to 
{\small{\begin{eqnarray}
 L_0^2 = \frac{2 g }{m \omega_z^2} \left\{ n_0({\bf 0})  +\frac{2}{ \lambda_{\rm
T}^3}\left( g_{3/2}\left[e^{-\frac{g n_0({\bf 0})}{k_{\rm B} T}}\right] - g_{3/2}\left[ 1 \right] \right) \right\}.
\end{eqnarray}}}
This compression effect is directly observed in
section \ref{expansion}, and indirectly through the measurements of condensed fraction in
section \ref{sectionfc}.

A further step forward would be to include collective effects in
the model. However, reference \cite{dalfovo1997a} points out that
low-energy, collective excitations cause a minute change in the
thermodynamic properties of the system even at relatively low
temperatures, $T \lesssim \mu$. Another approach, based on Quantum
Monte-Carlo calculations \cite{krauth1996a,holzmann1999a}, has
confirmed that the HF approximation could reproduce the {\it
thermodynamics} of the trapped clouds to a very good accuracy.


\begin{thebibliography}{37}
\bibitem{notetemp}
A possible cause for this behavior is that for our parameters, the
radial expansion is already asymptotic ($\omega_\perp t \approx
57.0$), whereas the axial expansion is not ($\omega_z t \approx
1.2$). Thus, the spatial distribution along $z$ might be
intermediate between the in-trap distribution and the asympotic
one. Note however that using the temperature measured along the
axial direction (with the same exclusion region than the one
chosen to find $T_x$) would merely shift the temperature downwards
by approximately $5-10\%$, without significantly affecting the
main conclusions of the paper.
\bibitem{visibhole}
Note also that integration along the probe line-of sight decreases
the visibility of the hole, even in the trap.
\end{thebibliography}

%


\begin{figure}[ht!]
\includegraphics[width=7cm]{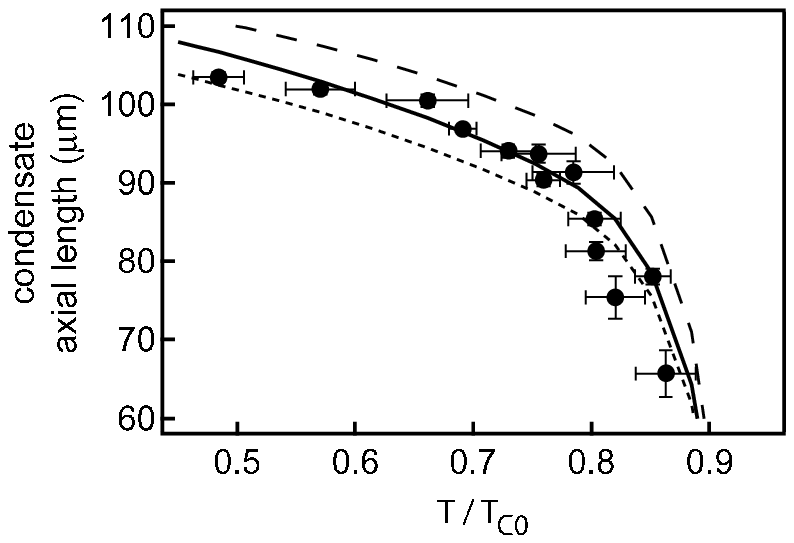}
\caption{Compression of the condensate by the thermal cloud. The
axial length of the condensate as measured in the absorption
images is shown, with the TF (dashed) and the HF predictions for
the trapped condensate (dotted) and after rescaling by the same
factor as a TF condensate (solid). } \label{becL}
\end{figure}

\end{document}